\documentclass{article}
\usepackage{amsfonts}
\usepackage{amssymb}
\usepackage{graphicx}
\usepackage{amsmath}

\input{tcilatex}

\begin{document}

%%%%%%%%%%%%%%%%%%%%%%%%%%%%%%%%%%%%%%%%%%%%%%%%%%%%%%%%%%%%%%%%%%%%%%%%

\title{Fisher zeros in the Kallen-Lehmann \\ approach to 3D Ising model.}
\author{Marco Astorino$^{1,2}$ \thanks{marco.astorino AT gmail.com}  , Fabrizio Canfora$^{2}$  \thanks{canfora AT cecs.cl} , Gaston Giribet$^{3}$ \thanks{ gaston@df.uba.ar} \\
%EndAName
\\
{\small $^{1}$\textit{Instituto de F\'{\i}sica, Pontificia Universidad Cat\'olica de Valpara\'{\i}so, Chile}} \\
{\small $^{2}$\textit{Centro de Estudios Cient\'{\i}ficos} (\textit{CECS), Valdivia, Chile}} \\
{\small $^{3}$\textit{Departamento de F\'{\i}sica, Universidad de Buenos Aires (UBA)}} \\ {\small \textit{and CONICET, Argentina}} }
\maketitle

\begin{abstract}

The distribution of the Fisher zeros in the Kallen-Lehman
approach to three-dimensional Ising model is studied.
It is argued that the presence of a non-trivial angle (a cusp) in
the distribution of zeros in the complex temperatures plane near
the physical singularity is realized through a strong breaking of
the 2D Ising self-duality. Remarkably, the realization of
the cusp in the Fisher distribution ultimately leads to an
improvement of the results of the
Kallen-Lehmann ansatz. In
fact, excellent agreement with Monte Carlo predictions both at
high and at low temperatures is observed. Besides, agreement between both approaches is found for the predictions of
the critical exponent $\alpha $ and of the universal amplitude
ratio $\Delta =$A$_{+}$/A$_{-}$, within the
3.5\% and 7\% of the Monte Carlo predictions, respectively.

\[
\]



CECS-PHY-08/15

\end{abstract}

\vspace{2 cm}

\section*{Introduction}

A proper understanding of non-perturbative phenomena in field
theory and statistical mechanics is a major challenge in
theoretical physics. Among the most important examples of such
phenomena it counts the problem of color confinement in QCD,
which, despite tireless attempts along the years, is still
begging for a theoretical description. Another renowned unsolved
problem is that of the Ising model in three dimensions. Actually,
both problems are known to be connected, as the Svetitsky-Yaffe
conjecture \cite{SV82} states that the three-dimensional Ising
model is closely related to the problem of color confinement and
that it likely captures its main non-perturbative features near
the transition. More generally, three-dimensional Ising model is
closely related to a large class of physical systems near the
critical point. Consequently, a deeper understanding of the
non-perturbative dynamics of the three-dimensional Ising model
would be of great importance in several areas of theoretical
physics \cite{PV02}. In turn, trying to find new
semi-analytical methods, non-perturbative in nature, to shed new light
on this problem represents a very interesting program.

In a recent work \cite{CPV06}, it has been proposed to address the problem of three-dimensional Ising model by using a method originally inspired in 
the spirit of Regge's theory of scattering \cite{Re1, Re2}. Regge theory is a fully non-perturbative approach which allows the description of many 
experimental data in terms of an ansatz with few parameters, these to be determined in comparison with observations and/or by using theoretical 
arguments. The idea in \cite{CPV06} was to propose an ansatz, also with few parameters, to describe the free energy of the three-dimensional Ising 
model. The ansatz is heuristically motivated by mimicking the relation existing between the expression for the free energy in one and two 
dimensions. Then, tuning the parameters of the model for it to describe the high temperature regime, one ends up with an expression that also 
reproduces the results at low temperature with remarkable accuracy.

This idea was first discussed in \cite{Ca07} and it was partially confirmed by the results of \cite{ACMP08},
where, even in its most simplified formulation (in which a {\it minimal
duality breaking} was assumed), such a method was shown to give results in a
surprising agreement with observations and Monte Carlo data.
In order to investigate the model in more detail (and
with modest computational resources) it is necessary to find a
theoretical tool able to fix (or at least to find some bounds on)
the {\it Regge parameters} appearing in the model of \cite{Ca07}. Here,
we report a remarkable progress in this direction.

A suitable tool to theoretically constraint the ansatz of
\cite{Ca07} is the analysis of the Fisher zeros, which
is a powerful technique developed by M. Fisher in the 1960's. Following the ideas of Yang and Lee \cite{LY}, Fisher
suggested to think of the inverse temperature $\beta $ as a
complex variable \cite{Fi68}. This technique permits to get relevant physical
information about the statistical system by looking at the analytic properties of the (extended) thermodynamical functions. For instance, the
distribution of zeros of the partition function in the complex
$\beta $ plane (Fisher zeros) near the critical point permits to
determine the universal amplitude ratio $\Delta = A_{+}/A_{-}$ of
the specific heat; see the classical papers \cite{Suzuki, Abe}. It has been also stressed that Fisher zeros technique
turns out to be useful in analyzing the strength of the phase
transitions (see for instance \cite{Kenna}\ \cite{Kenna2}\ and
references therein).

In this paper, we will be concerned with the analysis of Fisher
zeros within the framework of \cite{Ca07,ACMP08}. As it will be
explained in more details in the next sections, the ansatz
proposed in \cite{Ca07} appears to be well suited for this
kind of analysis since it is, in a sense, already in a {\it Fisher
zeros form}. This method suggests a simple way to realize a {\it
strong breaking of duality}, leading to an improvement of the
semi-analytical results in comparison with the numerical data.

The paper is organized as follows. In Section 1, a short
introduction to Fisher zeros is given. In Section 2, the relations
between Fisher zeros and duality breaking in Ising model is
discussed. In Section 3, the nice interplay between Fisher zeros
and the Kallen-Lehmann representation (KL) is analyzed in detail.
In Section 4, we introduce what we call the {\it strong duality
breaking} in the KL free energy for the three-dimensional Ising
model. In Section 5, it is shown how the strong duality breaking
improves the comparison with Monte Carlo results, both at high and
at low temperatures. In Section 6, the behavior of the critical
exponent at the critical point is described, together with the
behavior of the universal amplitude ratio of the specific heat
$\Delta $. Besides, some feasible further improvements are pointed
out. We conclude with a discussion in Section 7.

\section{Fisher zeros: a short introduction}

As it has been argued originally by M. Fisher \cite{Fi68}, a very
powerful method to analyze the thermodynamics of a generic spin
system is to complexify its temperature and to study the
distribution of zeros of the partition function in the complex
$\beta $ plane. Let $Z(\beta )$ and $F(\beta )$ be the partition
function and the free energy of the system under analysis,
respectively, and let $g(r)$ be the distribution of zeros of
$Z(\beta )$ in the complex $\beta $-plane, such that $g(r)dr$ is the number of zeros
of $Z(\beta )$ between $r$ and $r+dr$, where near the critical
point $\beta _c$ one has $\beta _{\mathbb{C}}=\beta _{c}+r\exp
(i\phi )$ (the meaning of the parameter $\phi $ is clarified
below). Then, one can show that, near the critical point
$\beta _{c} $, the non-analytic part of the internal energy (that
is, the singular part of the derivative of the free energy) and of
the free energy itself, responsible for the phase transition, have
the following form
\begin{eqnarray}
E(\beta ) &\approx &\partial _{\beta }F\approx \int\limits_{0}^{R}\frac{%
g(r)dr}{\beta _{c}-\beta +r\exp (i\phi )}+c.c.  \label{fi1} \\
F &\approx &\int\limits_{0}^{R}g(r)\log \left( \beta _{c}-\beta +r\exp
(i\phi )\right) dr  \nonumber % \label{fi1.25}
\end{eqnarray}%
where "$c.c.$" means the complex conjugate and $R$ is a suitable cut-off.
The angle $\phi $ is the angle formed by the tangent to the curve of zeros of $Z(\beta )$ and the real axis at the critical point $\beta _{c}$.
From the formulas above, one can derive an expression for the
singular part of the specific heat\footnote{here we will only consider the
case of second order phase transitions}
\begin{equation*}
C=-\beta _{c}^{2}\left( \int\limits_{0}^{R}\frac{g(r)dr}{\left( \beta
_{c}-\beta +r\exp (i\phi )\right) ^{2}}+c.c.\right) % \label{fi1.5}
\end{equation*}

Then, let us consider a system undergoing a second order phase
transition, so that the
singular part of the specific heat reads%
\begin{eqnarray*}
C &=&A_{+}\left\vert \beta -\beta _{c}\right\vert ^{-\alpha }\ \ \ \ \beta
<\beta _{c},\ \ \alpha >0, \\
C &=&A_{-}\left\vert \beta -\beta _{c}\right\vert ^{-\alpha }\ \ \ \ \beta
>\beta _{c}; \\
C &=&A_{+}\log \left\vert \beta -\beta _{c}\right\vert \ \ \ \ \beta <\beta
_{c},\ \ \alpha =0, \\
C &=&A_{-}\log \left\vert \beta -\beta _{c}\right\vert \ \ \ \ \beta >\beta
_{c},
\end{eqnarray*}%
where $\alpha $ is the critical exponent and, in general,
$A_{+}\neq A_{-}$. It is worth mentioning that the ratio $\Delta
=A_{+}/A_{-}$ is a universal quantity,
so that its computation is a relevant (and challenging) question. The ratio $%
\Delta $ can be measured in many interesting physical systems. It
can be shown (see, for instance, \cite{Suzuki} \cite{Abe}) that
both $g(r)$ near the critical point and the value of $\Delta $ are
related to $\phi $ in such a way that, for $\alpha \neq 0$, the
relations read
\begin{eqnarray*}
\Delta =\frac{\cos [(2-\alpha )\phi ]}{\cos [(2-\alpha )\phi +\alpha \pi ]} , \ \ \ \ 
g(r) \approx g_{0}r^{1-\alpha }
\end{eqnarray*}
so that the amplitude ratio is trivial (namely $\Delta =1$) when
the intersection between the zeros and the real axis is vertical
(corresponding to $\phi =\pi /2$). In the exact solution of the
Ising model in two dimensions \cite{O44} $\Delta =1$, but it is
not the case in three dimensions, where it is known that
\begin{equation*}
\Delta _{3D}\approx 0.55 \ , \ \ \ \  \phi _{3D}\neq \pi /2.
\end{equation*}
(see, for an updated review, \cite{PV02}.)

The aim of the present paper is to discuss the relation between
the existence of a cusp in the Fisher curve (i.e. the fact that
$\Delta \neq 1$) and a breaking of the 2D Ising self-duality. Such a duality symmetry is absent in the 3D Ising Model; consequently, any suitable 
attempt to dimensionally extend analytic methods from two to three dimensions would require to specify a precise mechanism of self-duality breaking. In 
\cite{ACMP08} 
self-duality was broken in the softest possible way. Here we will show that a stronger breaking is favoured by the comparison with 
observations.

\section{Fisher zeros and duality breaking}

Let us shortly describe the case of the two-dimensional
ferromagnetic\ Ising model. The free energy is given by
\begin{eqnarray}
F_{2D}(\beta ) &=&\log 2\cosh 2\beta +\frac{1}{2\pi }\int\limits_{0}^{\pi
}dt\log \left\{ \frac{1}{2}\left[ 1+\sqrt{1-k_{2D}(\beta )^{2}\sin ^{2}t}%
\right] \right\}   \label{fi2} \\
\left( k_{2D}(\beta )\right) ^{2} &=&\left( \frac{2}{\cosh 2\beta \coth
2\beta }\right) ^{2}=\left( 4\frac{\exp (2\beta )-\exp (-2\beta )}{\left(
\exp (2\beta )+\exp (-2\beta )\right) ^{2}}\right) ^{2},  \label{fi3} \\
0 &\leq &\left( k_{2D}(\beta )\right) ^{2}\leq 1.  \notag
\end{eqnarray}%

The critical point of the two-dimensional Ising model is located
at the maximum of $k_{2D}(\beta )$, namely
\begin{equation*}
\left( k_{2D}(\beta _{c})\right) ^{2}=1 % \label{fi4}
\end{equation*}%
and, \textit{being} $k_{2D}$ a smooth function, it is also true
that
\begin{equation*}
\left. \partial _{\beta }k_{2D}\right\vert _{\beta =\beta _{c}}=0.
\end{equation*}%

The function $k_{2D}$ (a sort of effective coupling constant)
encodes the duality properties of the model: if one writes
$k_{2D}$ in terms of $\tau =\tanh \beta $ (the high temperatures
variable) or in terms of $u=\exp (-2\beta )$ (the variable at
low temperatures) it turns out that $k_{2D}$ looks the same, and this is a
convenient way to express the self-duality of the 2D Ising model
(the self-duality of the 2D Ising model was discovered in
\cite{KW41}, before the discovery of its exact solution in 1944).
From the exact solution (\ref{fi1})-(\ref{fi2}) one can determine the Fisher zeros of the 2D Ising model, which lay on a circle in
the complex $u=\exp (-2\beta )$
plane. The equation determining the curve on which the Fisher zeros lay is:%
\begin{equation*}
\left\vert \left( k_{2D}(\beta )\right) ^{2}\right\vert =1  %\label{fi4.5}
\end{equation*}%
where $\beta $ is now a complex variable; while in the the complex $u$-plane
Fisher zeros have the following form:
\begin{equation*}
u=1-\sqrt{2}\exp (i\theta ). % \label{fi4.75}
\end{equation*}%

Being the Fisher zeros located on a circle whose center is on the
real axis, the intersection of the Fisher zeros with the real axis
is vertical and thus the angle $\phi $ is $\pi /2$, implying $\Delta
=1$. In the two-dimensional Ising model, the reason why the
intersection at the critical point is vertical is simple: the physical
singularity of the two-dimensional Ising model is the real
positive solution (in $\beta $) of the equation
\begin{equation}
1-\left( k_{2D}(\beta )\right) ^{2}\left( \sin t\right) ^{2}=0
\label{fi4.95}
\end{equation}%
which cannot be fulfilled unless
\begin{equation*}
\left( \sin t\right) ^{2}= \left( k_{2D}(\beta )\right) ^{2}=1.
\end{equation*}%

Then, let us call the complex zeros near the critical point $\beta
^{\ast }=\beta _{c}+x+iy$ (near the critical point $x$ and $y$ are
both small). Thus, $x$
is implicitly defined in terms of $y$ by the equation%
\begin{equation}
H_{2D}(x,y)=1-\left\vert k_{2D}(\beta _{c}+x+iy)^{2}\right\vert =0.
\label{fi5.5}
\end{equation}%

Near the critical point, when the intersection is not
vertical, one can think of $x$ as a monotonic function of $y$ (or
\textit{viceversa}). Therefore, in order for $x$ to be defined
implicitly in terms of $y$ by the equation above near $x=y=0$, the
derivative of $H_{2D}(x,y)$\ with respect to $y$ should be
different from zero. On the other hand, in order to have only one
physical singularity, one has to demand the
existence of only one solution of Eq. (\ref{fi4.95}) on the real positive $%
\beta $-axis and such a solution necessarily occurs at the maximum of $%
k_{2D}^{2}$.

As long as $k_{2D}^{2}(\beta )$ is a smooth function that admits an analytic
extension near $\beta _{c}$, the derivative of $H_{2D}(x,y)$ (defined as in Eq. (\ref{fi5.5})) 
with respect to $y$ vanishes at $x=y=0$, and this represents an obstruction to think of $x$ as
implicitly defined in terms of $y$ through (\ref{fi5.5}) at that point. So, no cusp in the Fisher curve could exist since the curve of distributions of zeros intersects the $x$-axis vertically.

On the other hand, in physics one is often interested in systems with
non-trivial ratio $\Delta \neq 1$, where the cusp manifests
itself. For instance, this is what happens in the Ising model in
three dimensions, in what we are interested here. In order to
analyze the possibility of achieving a realization yielding
$\Delta \neq 1$, first we have to make some comments on the
general form of the partition function:
let us assume for a moment that for a systems with only one phase transition, near
the critical point, the equation that determines the singularities
can be written in the form
\begin{equation}
1-\left( k_{eff}(\beta )\right) ^{2}\left( \sin t\right) ^{2}=0  \label{fi6}
\end{equation}%
for some real function $k_{eff}(\beta )$ whose only positive
maximum occurs at the critical point $\beta _{c}$, being $t$ some
dummy integration variable as in (\ref{fi1}). In such a case,
there is only one positive real solution occurring when both
$\left( k_{eff}(\beta )\right) ^{2}$\ and $\left( \sin t\right)
^{2}$\ are\ at their maxima (equal to one).

Thus, Fisher zeros can be described by an equation which is
formally analogous to the one of the two-dimensional Ising model;
namely
\begin{equation}
1=\left\vert \left( k_{eff}(\beta )\right) ^{2}\right\vert .  \label{fi6.5}
\end{equation}
This is part of the proposal in \cite{Ca07}. Actually one may introduce at least implicitly a suitable $k_{eff}$ for many statistical system (with standard Fisher zeros) in such a way that (\ref{fi6.5}) represents precisely the Fisher zeros of the system of interest. 

Equations (\ref{fi6}) and (\ref{fi6.5}) simply encode the fact
that there is only one physical singularity. Function
$k_{eff}(\beta )$ indicates how far from being self-dual a system
is. Therefore, according the discussion above, having a value $\Delta \neq 1$ would imply $k_{eff}(\beta
)$ in (\ref{fi6}) to be non-differentiable at the critical point,
where it takes its maximum value. Actually, in order to achieve a cusp in the distribution of the Fisher zeros, it is enough to require a non-analiticity of $k_{eff}(\beta )$ at the critical point.  As a matter of fact, our results indicate that the first derivative of $k_{eff}(\beta )$ is continuous while the second derivative is discontinuous. This is one of the
hints for constructing our ansatz, as we are reminded of the fact that in three dimensions $\Delta _{3D} <1$.

In \cite{ACMP08}, for the sake of simplicity (and because of the
computational resources) a minimal duality breaking was assumed,
according to what $k_{eff}(\beta )$ still was smooth at the
critical point. As a result, the agreement with Monte Carlo data
and observations was found to be very good. Nevertheless, as it
will be explained in the next sections, when the cusp ($\Delta
\neq 1$) in the Fisher curve is implemented, the comparison with
Monte Carlo data turns out to be even better, showing a new and more efficient way of
exploring the space of parameters of the model.

\section{Fisher zeros in the Kallen-Lehmann representation}

A powerful non-perturbative technique in field theory is the Kallen-Lehmann representation.
Such a tool allows to encode in a very natural way many analytical properties of the spectral functions. It is interesting to adapt this method (often 
exclusively associated to Quantum Field Theory) to the analysis of the 3D Ising model. 
This representation \cite{Ca07} gives rise to an ansatz for the
free energy of the three-dimensional Ising model of the following form\footnote{%
The "Regge parameters" appearing in these formulas are related with those
appearing in \cite{Ca07} by the following identities $\zeta _{1}=\nu $, $%
\zeta _{3}=1/2=\zeta _{2}$, $\zeta _{0}=\alpha $.}
\begin{eqnarray}
\label{parto32}
F_{3D}^{(\zeta _{i},\lambda )}(\beta ) &=&F_{2D}(\beta )+\frac{\lambda }{%
\left( 2\pi \right) ^{2}}\int\limits_{0}^{\pi }dz\int\limits_{0}^{\pi
}dy\cdot  \notag \\
&&\cdot \log \left\{ \frac{1}{2}\left[ 1+\left( 1-\left[ 2\frac{\left(
\Delta (z)-1\right) ^{\zeta _{1}}}{\Delta (z)}\right] ^{\zeta _{2}}\sin
^{2}y\right) ^{\zeta _{3}}\right] \right\} ,  \label{fi7}
\end{eqnarray}
where
\begin{eqnarray}
\ \Delta (z) &=&\left( 1+\left( 1-k_{eff}(\beta )^{2}\sin ^{2}z\right)
^{\zeta _{0}}\right) ^{2},\ \zeta _{0},\zeta _{1},\zeta _{2},\zeta _{3}>0,
\label{fi7.25} \\
0 &\leq &\left( k_{eff}(\beta )\right) ^{2}\leq 1,\ \ \ 1\leq \Delta (z)\leq
4,  \label{fi7.5}
\end{eqnarray}
and where the values of the parameters $  \zeta _{0},\zeta _{1},\zeta _{2},\zeta _{3}  $ (henceforth {\it Regge Parameters}) in the case of the two-dimensional
Ising model would correspond to $\lambda =1$, $\zeta _{i}=1/2$ for $i=0,..,3$.

Expression (\ref{fi7}) will be referred as Kallen-Lehmann free energy. The heuristic argument that suggests to write down (\ref{fi7}) goes as follows:
think of an operator $\mathbf{O}%
_{1\rightarrow 2}$ which "dresses" the trivial one-dimensional
solution of the Ising model giving rise to the Onsager solution.
Then, the Kallen-Lehmann ansatz for the free-energy for the three-dimensional
Ising model is obtained by
modifying $\mathbf{O}_{1\rightarrow 2}$ in such a way that the parameters%
\footnote{Namely, in the two-dimensional Ising model, the parameters which appear in $%
\mathbf{O}_{1\rightarrow 2}$ are $\lambda =1$, $\zeta _{i}=1/2$. In order to
change the critical exponents one needs to change (at least) $%
\lambda $ and $\zeta _{i}$.} $\lambda = 1$ and $\zeta _{i}=1/2$
now become free parameters, and then applying such a modified
dressing operator $\mathbf{O}_{2\rightarrow 3}$ to the Onsager
solution. The result is Eq .(\ref{fi7}).

A nice feature of this dimensional-recursive method is that it permits to distinguish
between the problem of fixing five external Regge parameters\footnote{In \cite{ACMP08}, for sake of simplicity, the parameters $\zeta _{1}$ and $%
\zeta _{3}$ were fixed to 1/2; that is, to the values they take in the two-dimensional problem. Thus, already
by using only three parameters, very good results in comparisons
with Monte-Carlo data were obtained.} $\lambda $ and $\zeta _{i}$
($i=0,..,3$), and the problem of determining the function $k_{eff}(\beta
)$. The function $k_{eff}(\beta )$ plays the same role that
$k_{2D}(\beta )$ plays in (\ref{fi3}). Namely, it encodes the
information about how different the degrees of freedom are in the
low and the high temperature regimes. Thus, at least in principle,
one could fix $k_{eff}(\beta )$ by analyzing how the duality is
broken in the three-dimensional Ising model. Then, Regge
parameters can be found by looking, for instance, at the high
temperatures behavior.

An important piece of information is given by knowing which are
the more relevant parameters in writing down the ansatz
(\ref{fi7}), as well as understanding their physical
interpretation. In particular, as it will be discussed in the next
sections, one learns that tuning appropriately just three Regge
parameters is sufficient to get an excellent agreement, both at
high and at low temperatures. Besides, a very good description
near the critical point is also observed. Our
results suggest that the agreement with observations and Monte Carlo data
improves its performance by resorting to a careful analysis of duality breaking, instead of 
trying to select the Regge parameters {\it ad hoc}. From our analysis herein we achieve a substantial
improvement of the method of \cite{Ca07}.

Nevertheless, it is worth mentioning that our analysis is still far from
being an exhaustive exploration of the whole space of parameters
of the model. In fact, this is what makes the results more
surprising, as the agreement with Monte Carlo data is excellent
despite only a small piece of the moduli space was analyzed.

Let us now compute the internal energy in the Kallen-Lehman
representation. A simple computation yields
\begin{eqnarray}
E_{3D}^{(\zeta _{i},\lambda )}(\beta ) &\approx &\partial _{\beta
}F_{3D}^{(\zeta _{i},\lambda )}(\beta )=E_{2D}(\beta )+\frac{\lambda }{%
\left( 2\pi \right) ^{2}}\int\limits_{0}^{\pi }dz\int\limits_{0}^{\pi
}dy\left\{ \left( -\frac{\zeta _{3}\zeta _{2}\sin ^{2}y}{N}\right)
\right. \times  \notag \\
&& \times  \left. D^{\zeta _{3}-1}R^{\zeta _{2}-1}(\partial _{\Delta }R)\frac{%
\partial \Delta }{\partial \left( k_{eff}(\beta )^{2}\right) }\frac{\partial
\left( k_{eff}(\beta )^{2}\right) }{\partial \beta }\right\} ,
\label{fi9}
\end{eqnarray}
where
\begin{eqnarray}
N =1+\left( 1-\left[ 2\frac{\left( \Delta (z)-1\right) ^{\zeta _{1}}}{\Delta (z)}\right] ^{\zeta _{2}}\sin ^{2}y\right) ^{\zeta _{3}},\ \ N\geq 1, \nonumber \\
%\label{fi10} \\
D =1- R^{\zeta _{2}}\sin ^{2}y, \ \ \ R=2\frac{\left( \Delta (z)-1\right) ^{\zeta _{1}}}{\Delta (z)},
\label{fi11}
\end{eqnarray}
and
\begin{eqnarray*}
\frac{\partial \Delta }{\partial \left( k_{eff}(\beta
)^{2}\right) } &=&-2\left( 1+\left( 1-k_{eff}(\beta )^{2}\sin
^{2}z\right) ^{\zeta _{0}}\right) \left( \zeta _{0}\sin
^{2}z\right) \left( 1-k_{eff}(\beta
)^{2}\sin ^{2}z\right) ^{\zeta _{0}-1}, \nonumber \\ %\label{fi12} \\
\left( \Delta (z)-1\right) ^{\zeta _{1}} &=&\left\{ \left[ 2+\left(
1-k_{eff}(\beta )^{2}\sin ^{2}z\right) ^{\zeta _{0}}\right] \left[ \left(
1-k_{eff}(\beta )^{2}\sin ^{2}z\right) ^{\zeta _{0}}\right] \right\} ^{\zeta
_{1}}, \nonumber	\\  % \label{fi13} \\
\partial _{\Delta }R &=&2\frac{\left( \Delta (z)-1\right) ^{\zeta _{1}-1}}{%
\left( \Delta (z)\right) ^{2}}\left[ 1+\left( \zeta _{1}-1\right) \Delta (z)%
\right]  \label{fi14}
\end{eqnarray*}

It is worth noticing that, from the point of view of Fisher
zeros distribution, there is a special value for $\zeta _{1}$. In order to have
a "singularity equation" similar to the one arising in the
two-dimensional Ising model, it is necessary to ask that all the
singular terms of the internal energy have their origin in factors
like
\begin{equation}
\left( 1-k_{eff}(\beta )^{2}\sin ^{2}z\right) ^{e_{i}}  \label{fi15}
\end{equation}
where the exponents $e_{i}$ depend on the Regge parameters $\zeta
_{i}$. A potentially disturbing term in (\ref{fi9}), which is
not of the form (\ref{fi15}), is $D^{\zeta _{3}-1}$, since $D=1-R^{\zeta _{2}}\sin ^{2}y$ could be negative and $\zeta _{3}-1$\ could be 
negative\footnote{This term is also disturbing since it could give rise to complex
numbers in the case $D$ is negative and $\zeta _{3}-1$\ is
negative.}. Then, to avoid undesired divergences in the internal energy,
one may ask whether a special of $\zeta _{1}$\ exists such that the maximum\footnote{Where one "maximizes" $C$ over the $\Delta $ fulfilling Eq. (\ref{fi7.5}).} $%
m_{0}$ of $R$
\begin{equation*}
m_{0}\underset{\Delta }{=\max }R\underset{\Delta }{=\max }\left\{ \left[ 2\frac{\left( \Delta -1\right) ^{\varsigma _{1}}}{\Delta }\right] \right\} ,
\end{equation*}
is less than or equal to one, and such that $m_{0}$ does not
depend on $z$ and $\beta $:
\begin{equation*}
0\leq \left[ 2\frac{\left( \Delta -1\right) ^{\varsigma _{1}}}{\Delta }%
\right] \leq m_{0}=1. % \label{lowte2}
\end{equation*}%
Interestingly enough, such a special value of $\varsigma _{1}$ does exist and it is precisely $1/2$, the value of the
two-dimensional Ising model. Thus, with $\varsigma _{1} = 1/2$, when function $R$ in (\ref{fi11}) attains its maximum $m_{0}$ (this
happens for $\Delta =2$), it is always multiplied by its derivative $%
\partial _{\Delta }R$ (see Eq. (\ref{fi14})), which vanishes. In this way, all the potential singular
terms in Eq. (\ref{fi9}) are of the {\it Fisher form} in Eq.
(\ref{fi15}) so that they can be analyzed with the method
described in the previous sections. The condition $\varsigma_{1}=1/2$ (which in \cite{ACMP08} was imposed for
simplicity) is obtained here as being the appropriately value from the
point of view of the Fisher zeros.
Thus, the Kallen-Lehmann free energy is a smooth function everywhere apart from the critical point singled out by the equation (\ref{fi6.5}).

\section{Duality breaking and $k_{eff}(\beta )$}

The previous analysis, relating Fisher zeros to duality breaking, suggests how to break duality taking into account
the non-triviality of the amplitude ratio $\Delta \neq 1$ in three
dimensions. A second important ingredient in the discussion is the
Marchesini-Shrock symmetry \cite{MS88}, which states that the
partition function of Ising model on regular
hypercubic lattices is invariant under
\begin{equation*}
\beta \rightarrow \beta +in\frac{\pi }{2},\ \ n\in
%TCIMACRO{\U{2124} }%
%BeginExpansion
\mathbb{Z}
%EndExpansion
.
\end{equation*}
The simplest possible choice of $k_{eff}(\beta )$ which satisfies
this constraint is
\begin{equation*}
k_{eff}(\beta )=4\frac{d_{3}\exp (2\beta )-d_{2}\exp (-2\beta )}{\left(
d_{1}\exp (2\beta )+d_{0}\exp (-2\beta )\right) ^{2}}. % \label{kaeff}
\end{equation*}

In \cite{ACMP08}, two of the parameters $d_{i}$ ($i=0,..,3$) have
been fixed by asking the expected transition to occur at
\begin{equation}
k_{eff}(\beta _{c})=1,\ \ \ \ \beta _{c}=0.22165,  \label{co2}
\end{equation}
and by requiring the vanishing of the derivative of $k_{eff}(\beta)$ in $\beta _{c}$, namely
\begin{equation}
\left. \beta ^{\ast }\right\vert \partial _{\beta}k_{eff}(\beta )\left\vert _{\beta =\beta ^{\ast }}=0 \right.
\label{co12}
\end{equation}

On the other hand, the other two parameters $d_i$ were chosen to be equal to those values they take in the two-dimensional model. However, from
the previous discussion, we learn that condition (\ref{co12})
should not be kept in the cases of systems exhibiting non-trivial
amplitude ratio $\Delta $. Thus, we will assume that $\beta _{c}$
is a maximum of $k_{eff}$\ but with a left-derivative different
from the right-derivative at $\beta _{c}$.

Indeed, there are an infinite number of possible choices
of parameters that leads to strong duality breaking such that
$k_{eff}(\beta )$ has a cusp at the critical point. As in
\cite{ACMP08}, we will follow a simplicity criterion: we will assume that both for $\beta <\beta _{c}$ and for $\beta >\beta _{c}$ the
function $k_{eff}(\beta )$ has a form similar to that of
$k_{2D}(\beta )$ in two dimensions (the numerical results indicate that the discontinuity appears in the second derivative, while the first derivative at the critical point is continuous). In addition, we will consider two constraints on the parameters: the first is
that, for $\beta <\beta _{c}$, $k_{eff}(\beta )$ is decreasing,
while for $\beta <\beta _{c}$\ it is increasing; the second is the
continuity of $k_{eff}(\beta )$ together with Eq. (\ref{co2}).
Summarizing, we have
\begin{eqnarray}
k_{eff}(\beta ) &=&k_{eff}^{(+)}(\beta )=4\frac{d_{3}^{(+)}\exp
(2\beta )-d_{2}^{(+)}\exp (-2\beta )}{\left[(d_{1}^{(+)}\exp
(2\beta )+d_{0}^{(+)}\exp (-2\beta )\right] ^{2}}\ \ \ , \beta
<\beta _{c}
\label{fi16} \\
k_{eff}(\beta ) &=&k_{eff}^{(-)}(\beta )=4\frac{d_{3}^{(-)}\exp
(2\beta )-d_{2}^{(-)}\exp (-2\beta )}{\left[ d_{1}^{(-)}\exp
(2\beta )+d_{0}^{(-)}\exp (-2\beta )\right] ^{2}}\ \ \ , \beta
>\beta _{c}
\label{fi18}
\end{eqnarray}
obeying
\begin{eqnarray}
k_{eff}^{(+)}(\beta _{c}) &=&k_{eff}^{(-)}(\beta _{c})=1,  \label{fi19} 
\end{eqnarray}
and
\begin{eqnarray}
\partial _{\beta }k_{eff}(\beta ) &>&0\ \ , \ \beta <\beta _{c}, \label{fi20} \\
\partial _{\beta }k_{eff}(\beta ) &<&0\ \ , \ \beta >\beta _{c}.
\label{fi20.5}
\end{eqnarray}

Eq. (\ref{fi19}) fixes one of the $d_{i}^{(+)}$ (as well as one of
the $d_{i}^{(-)}$) in terms of the others and the
critical temperature, while Eqs. (\ref{fi20}) and (\ref%
{fi20.5}) ensure that $\beta _{c}$ is actually a maximum. In
principle, the relevant "duality breaking" parameter (to be found
by comparing the theory with the available data) is the
discontinuity $\Delta k'$ of the first derivative of $k_{eff}
(\beta )$, namely
\begin{equation}
\Delta k' = \left( \partial _{\beta }k_{eff}^{(+)}(\beta)-\partial
_{\beta }k_{eff}^{(-)}(\beta)\right) \Big|_{\beta=\beta_c}.  \label{sdbp}
\end{equation}

However, from the practical point of view, it is much easier to
work with the parameterization (\ref{fi16})-(\ref{fi18}),
where a non-vanishing $\Delta k'$ is seen to be given by a
non-vanishing $\Delta _{d_{i}} =d_{i}^{(+)}-d_{i}^{(-)} \neq 0,\ i=0,..,3 $, which implies $\Delta k'\neq 0$.

Thus, some of the parameters $d_{i}^{(+)}$ and $d_{j}^{(-)}$ can
be appropriately chosen in order to improve the agreement with the
{\it experiments}.

The parameterization (\ref{fi16}) and (\ref{fi18}) is actually
the simplest one being compatible with the {\it
strong duality breaking}, inspired by the exact solution
of the two-dimensional model. So, in principle, the results
obtained by this method could be improved if it were
possible to explore the whole space of parameters more
exhaustively.

Because of limitations in computational resources, we are able to
explore only in a very restrictive region of the space of parameters of
\cite{ACMP08}. Even in this case, the results are remarkably good
if we make use of some hints. We will proceed as follows: first,
we will chose some of the parameters to be equal to the best
values found in \cite{ACMP08}. Then, the novelty here will be to
introduce a parameter that controls the {\it strong duality
breaking}. In principle, one would expect the optimal set of parameters to be
far from the set found in \cite{ACMP08}, where quite simplifying hypothesis of minimal
duality breaking were assumed. However, as a very encouraging
signal, one finds that a very good agreement with observations is
found by studying this region of the moduli space. Thus, from now
on, we will fix
\begin{equation*}
\zeta _{0}=\zeta _{0}^{\ast }=1.9389,\ \ \ \zeta _{1}=\ \zeta _{1}^{\ast }=%
\frac{1}{2},\ \ \ \zeta _{2}=\zeta _{2}^{\ast }=1.9205, %  \label{mdb1}
\end{equation*}
and we will test the strong duality breaking in a small
neighborhood varying $\lambda $,$\ \zeta _{2}$ and $%
\Delta k'$.

Afterwards it will be useful to express $k_{eff}^{(-)}$ in terms of the low temperatures
variable $u$ as follows
\begin{equation}
k_{eff}^{(-)}(u)=4\frac{u\left( d_{3}^{(-)}-d_{2}^{(-)}u^{2}\right) }{\left(
d_{1}^{(-)}+d_{0}^{(-)}u^{2}\right) ^{2}}.  \label{filow1}
\end{equation}

Now, let us move on and study the high and low temperature regimes.

\section{High and low temperatures}

The idea of this section is to find the optimal set of parameters $\left(
\zeta _{3}^{\ast },\lambda ^{\ast },\Delta k'^{\ast }\right) $ in Eqs. (\ref{fi7.25}) and (\ref{sdbp}) that reproduce as close as possible
the available Monte Carlo data at high temperature (see \cite{AF02}). A hyper-cubic
lattice has been chosen in the parameters space (every point in
the lattice representing a
possible set of high temperature parameters), then the free energy (\ref{fi7}%
) will be evaluated at every point of the lattice. The optimal choice
of parameters will be the one minimizes the following deviation
function which, to some extent, represents the deviation between
the ansatz and the Monte Carlo data,
\begin{equation*}
\chi (\zeta ,\nu ,\lambda )=\sum\limits_{i}^{50}\left\vert F_{3D}^{(\zeta
,\nu ,\lambda )}(\beta _{i})+I_{0}-\left. F\right. _{HT}^{\ \ MC}(\beta
_{i})\right\vert ^{2},\  % \label{errhigh0}
\end{equation*}
where
\begin{equation*}
\beta _{i}-\beta _{i-1}=\frac{0.03}{50}\ \ ,\qquad \beta _{50}=\beta
_{max}=0.03,\ \ \ I_{0}=2.4819
\end{equation*}%
\begin{eqnarray*}
\left. F\right. _{HT}^{\ \ MC}(\beta ) &=&3\cosh \beta +\left( 3\left( \tanh
\beta \right) ^{4}+22\left( \tanh \beta \right) ^{6}+187.5\left( \tanh \beta
\right) ^{8}+\right. \\
&+&\left. 1980\left( \tanh \beta \right) ^{10}+24044\left( \tanh \beta
\right) ^{12}+319170\left( \tanh \beta \right) ^{14}+...\right)
\end{eqnarray*}%
Here, we keep the terms up to the $15^{th}$ order of \cite{AF02}
since our algorithm is not sensitive to higher order terms.
$\left. F\right. _{HT}^{\ \ MC}$ is the high temperature Monte
Carlo free energy, $\beta _{m}$ can be assumed to be of
order\footnote{Above $\beta \approx 0.05$ is not in
the high temperature regime as the critical temperature is at
$\beta ^{\ast }\approx 0.22$ which is only a factor of four
larger. Indeed, $\beta \approx 0.03$ appears to be not small
enough. Nevertheless, we will see that the agreement of our semi-analytical free energy with Monte Carlo data is excellent up to $\beta \approx
0.03$.} $0.03$, and $I_{0}$ is a constant introduced
for numerical convenience.

In order to compare the Regge coefficients with the high temperature
coefficients in \cite{AF02}, we need to change variable from $\beta $ to $%
t=\tanh \beta$.

Candidates to be the best parameters are
\begin{equation}
\zeta _{3}^{\ast }=0.3273,\ \ \ \  \frac{\lambda ^{\ast }}{(2\pi)^2}=0.1095 ,
\label{OHT}
\end{equation}
while the discontinuity at the critical point is given by $ \Delta k' = 0.75 \cdot 10^{-4} $ (which according to our precision is compatible with zero). As expected, the discontinuity of second derivative $ \Delta k''$ is of order one.
This can be achieved, for instance, by the following choice of $d_{i}^{(\pm)}$:
\begin{align*}
d_0^{(+)}&= 1 \ \ , \ d_1^{(+)} =  1  \ \ , \ d_2^{(+)}=-0.1575  \ \ , \ d_3^{(+)}= 0.7116   \\
d_0^{(-)}&= 0.27 \ \ , \ d_1^{(-)} =  1  \ \ , \ d_2^{(-)}= 0.3498  \ \ , \ d_3^{(-)}= 0.6251 
\end{align*}

The agreement appears to be excellent: the deviation at high temperatures turns out to be $\sigma _{HT}(\zeta ^{\ast },\nu ^{\ast },\lambda ^{\ast })\approx \sqrt{\chi (\zeta ^{\ast },\nu ^{\ast },\lambda ^{\ast })/50}\approx 3 \cdot 10^{-6}$, which is certainly compatible with Monte Carlo results (see for instance \cite{PV02} and references therein).

Once the optimal set of parameters have been found at high temperature,
it is observed that such set also leads to a good agreement at low
temperatures. The internal energy at low temperature is
($k_{eff}^{(-)}(\beta )$ has to be expressed in terms of $u$ as in
Eq. (\ref{filow1}))
\begin{equation}
\left\langle \frac{E}{N}\right\rangle ^{KL}(u)+2I_{1}=2u\frac{\partial }{%
\partial u}F_{3D}^{(\zeta _{3},\lambda ,\Delta k')}(u)\ .  \label{parto33}
\end{equation}

This is the average energy per spin\footnote{to be more precise, both expressions can
differ by a $I_{1}$.} to be
compared with $\left\langle \frac{E}{N}\right\rangle ^{MC}$, the
polynomial form in $u$ which represents the Monte Carlo average energy for
spin for small $u$ found in \cite{BCL92}.

We learned in \cite{ACMP08} that it is not convenient
to use the expression on the right hand side of (\ref{parto33}). 
Instead, the polynomial expression of \cite{BCL92} can be used to obtain the
Monte Carlo estimation for the free energy at low
temperatures; namely
\begin{equation*}
\left\langle \frac{E}{N}\right\rangle ^{KL}(u)+2I_{1}=2u\frac{\partial }{%
\partial u}F_{LT}^{\ \ \ MC}(u)=\sum_{i=6}^{14}a_{i}^{(L)}u^{i}
\end{equation*}
where
\begin{equation*}
F_{LT}^{\ \ \ MC}(u)=\frac{1}{2}\left( \sum_{i=6}^{14}\left( \frac{%
a_{i}^{(L)}}{i}\right) u^{i}-2I_{1}\log u\right) . % \label{lowFE}
\end{equation*}%
Low temperature test function reads
\begin{equation*}
\chi _{LT}(\zeta ^{\ast },\nu ^{\ast },\lambda ^{\ast
})=\sum_{i=1}^{50}\left\vert F_{3D}^{(\zeta ^{\ast },\nu ^{\ast },\lambda
^{\ast })}(u_{i})+I_{2}-F_{LT}^{\ \ \ MC}(u_{i})\right\vert ^{2},
%\label{comme2}
\end{equation*}%
\begin{equation*}
u_{i}-u_{i-1}=\frac{0.03}{50},\ \ \ \ u_{1}=u_{min},\ \ \ \ u_{50}=u_{\max
}=0.03,
\end{equation*}%
\begin{eqnarray*}
F_{LT}^{\ \ \ MC}(u) &=&\left[ \frac{1}{2}\left( \frac{12}{6}\left( u\right)
^{6}+\frac{60}{10}\left( u\right) ^{10}+\right. \right. \\
&&\left. \left. -\frac{84}{12}\left( u\right) ^{12}+\frac{420}{14}\left(
u\right) ^{14}...\right) +I_{1}\log u\right]
\end{eqnarray*}%
where $I_{1}=-1$ and $I_{2}=0.0954$. Our algorithm is sensitive up to the $15th$ order of the polynomial expression of
\cite{BCL92}, so we keep all these terms. $u_{\max }$ has to be much smaller than $u_{crit}=\exp (-2\beta _{crit})\approx 0.6$, so it can be reasonably 
assumed to be of order $0.03$.

The deviation at low temperature between the Kallen-Lehmann and the
Monte Carlo free energies \textit{evaluated for the same optimal parameters}
in (\ref{OHT}), which have been found by asking the optimal agreement at
high temperature, is $\sigma _{LT}(\zeta ^{\ast },\nu ^{\ast },\lambda ^{\ast })\approx \sqrt{\chi _{LT}(\zeta ^{\ast },\nu ^{\ast },\lambda ^{\ast })/50}%
\approx 7 \cdot 10^{-6}.$

Remarkably, one finds the agreement at low temperature to be an order of magnitude better than that in the case of minimal duality breaking studied in \cite{ACMP08}.

Moreover, in the next section we will show that, besides improving the agreement with Monte Carlo results at low and high temperature, the 
implementation of the strong duality breaking also permits to describe features at the critical point.

\section{The Critical Point}

Once the parameters have been fixed as in (\ref{OHT}) one may verify that the behavior at the critical point
is correctly reproduced as well. To do this one can fit near the critical point
the non-analytic\footnote{That is, one has to exclude the term $\log 2\cosh \beta $ which does not
contribute to the critical behavior.} part of the free energy in Eq. (\ref{parto32}) with the optimal parameters with a
function of the form
\begin{eqnarray*}
F_{CRIT} &\approx &A_{+}\left\vert \beta -\beta ^{\ast }\right\vert
^{2-\alpha }+c\ \ \ \ \beta <\beta _{c} \\ % \label{ficri1} \\
\label{critest} 
F_{CRIT} &\approx &A_{-}\left\vert \beta -\beta ^{\ast }\right\vert
^{2-\alpha }+c\ \ \ \ \beta >\beta _{c}  % \label{ficri2}
\end{eqnarray*}%
and find the optimal values of the constants $c$, $A_{+}$, $A_{-}$ and $\alpha $, so
that $\alpha $ will be our prediction for the critical exponent and $%
A_{+}/A_{-}$ will be our estimate for the universal amplitude ratio. This
form of the free energy's critical part is expected both from Conformal
Field Theory and from experiments. The results of the fit done\footnote{A regular sampling is chosen with step $ \Delta \beta = 2 * 10^{-5}  $, thus 
the points of the sampling of Kallen-Lehmann are of the form $\beta_{min}+ n \Delta \beta $ and $\beta_{max}- m \Delta \beta $ up to the critical point.} with \textit{Mathematica} by fitting the (non-analytic part of the) Kallen-Lehman
free energy with the optimal parameters in Eq. (\ref{OHT}) with the above
functions (\ref{critest}) from $\beta_{min} =0.22104$ to $\beta_{max} =0.22208$ yield
\begin{equation}
A_-=4.18, \ \ \  A_+=2.27 , \ \ \ c=-0.19,\ \ \ \alpha =0.11  \label{predi}
\end{equation}%
the agreement appears to be very good when compared with recent estimations
in \cite{BC02}, where the value $\alpha _{obs}\approx 0.114(6)$ was found. In turn, we find
\begin{eqnarray*}
\Delta \alpha &\approx &\frac{\alpha _{obs}-\alpha }{\alpha _{obs}}\approx  0.035  \\
\Delta (A_{+}/A_{-}) &\approx &\frac{({A_{+}}/{A_{-}})_{obs}-({A_{+}}/{A_{-}})}{({A_{+}}/{A_{-}})_{obs}}\approx 0.07
\end{eqnarray*}%

\begin{figure}[!htp]
\caption{ {\small Bold points in the graph represent a sampling of Kallen-Lehmann free energy $F_{3D}^{(\protect\zeta_i ^{\ast },\protect\lambda ^{\ast 
})}(\protect\beta)$, while the continuous line corresponds to the critical
part of the expected free energy $F_{CRIT}$ versus $\protect\beta$. One can observe the different leanings in the left and right sides of the critical 
point, which is ultimately related to a non-trivial $\Delta$.}}
\label{fig-cp}
\begin{center}
\includegraphics[angle=0, scale=0.21] {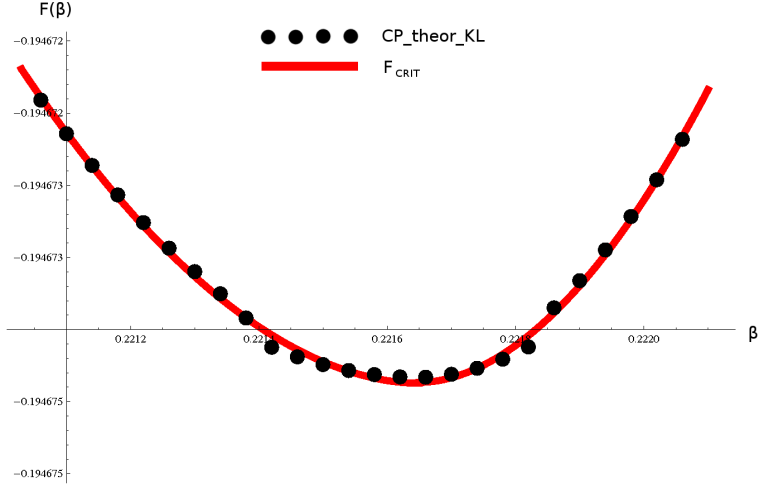}
\end{center}
\end{figure}

Both theoretical and experimental determinations of $\alpha $ can
be found in \cite{PV02}. The deviation $\Delta \alpha $ of our prediction (\ref{predi}) from observations appears to be less than 4\% in all the more
recent values, improving the minimal duality breaking results \cite{ACMP08}.
Figure (\ref{fig-cp}) shows the matching between the (non-analytic part of the) Kallen-Lehmann free energy and the $F_{CRIT}$, and it confirms that in 
the parts of the graph of $F_{CRIT}$ in which the
dependence on $\alpha $ is important the agreement is satisfactory.

We expect that more general strong duality breaking can improve the very
good agreement with observations and numerical data we found at high, low
temperatures as well as at the critical point. For sake of simplicity
(and because of our computational resources) we considered the simplest form
of strong duality breaking in which (inspired by the Onsager solution in two
dimensions) the functional form of $k_{eff}$ at high and at low temperature
is the same (see Eqs. (\ref{fi16}) and (\ref{fi18})). In turn, in our approach the cusp in the Fisher curve was implemented by the difference $\Delta _{d_{i}}$ between the coefficients $d_{i}^{(+)}$ at
high temperatures and the corresponding coefficients $d_{i}^{(-)}$ at low
temperature. However, one could argue that it is likely the case the functional form of $k_{eff}$\ at low
temperature in Eq. (\ref{fi18}) to be substantially different from the one at high
temperature in Eq. (\ref{fi16}). In fact, there are infinite possible terms
compatible with the Marchesini-Shrock symmetry that one can add to (\ref{fi18}) (still satisfying conditions (\ref{fi19}) and 
(\ref{fi20.5})) so that, in a sense, the very good agreement we found here is
unexpected. It would be very interesting to explore theoretical arguments
to further constrain $k_{eff}$. As a matter of fact, one may also try
to add terms to $k_{eff}$ compatible with the above physical requirements
and see what happens. However, without a theoretical guide, the time required for an exhaustive exploration of the moduli space would be out of 
range. The lack of sophisticated computational resources restricted us to explore a little region of the space of parameters. In particular, some 
parameters ($\zeta_0$, $\zeta_2$ , $\lambda$) were here kept fixed to take the same values found in \cite{ACMP08} just as a working 
hypothesis, and it is likely the case the optimal values to be slightly different from those of the ansatz in \cite{ACMP08}.
Nevertheless, the 
implementation of the strong duality breaking led us to a quality leap, bringing our predictions at low temperature closer to Monte Carlo 
simulations. Remarkably, at the 
same time our model permits to reproduce the universal amplitude ratio $\Delta$ and to predict the critical exponent $\alpha $.

%An attractive feature of our model that even just fixing the
%first few coefficients at high temperatures one obtains a good agreement at the critical point and at low temperatures.
%This permit to bypass the stiff and problematic numerical issue of the higher order numerical derivatives (as explained in \cite{ACMP08}) which seems 
%not to be deecisive in the critical point description as in other approches (see \cite{PV02} and references therein).

\section{Conclusions}

In this paper we studied the distribution of Fisher zeros in the Kallen-Lehmann approach to the three-dimensional Ising model. It was shown 
that non-trivial 
amplitude ratios
$\Delta = A_+ / A_-$ are compatible with the ansatz proposed in \cite{CPV06}. We proposed a mechanism (which we called {\it strong 
duality 
breaking}) by generating a cusp in the curve of the zeros of the 
free energy $F(\beta )$ in the complex $\beta $-plane, being the cusp located at the critical point.

This mechanism not only permitted us to reproduce the Monte Carlo prediction for the value of $\Delta $, but also led us to improve the results of 
\cite{ACMP08} bringing the predictions of the Kallen-Lehmann ansatz at low temperature closer to Monte Carlo simulations. The agreement turns out to be 
remarkable. 
In particular, the matching between our 
results and Monte Carlo estimations for the critical exponent $\alpha $ and for the amplitude ratio $\Delta $ exhibit a relative deviation of 3.5\% 
and 7\% respectively. 

We interpret our result as a motivation to continue the investigation of phenomenological semi-analytic methods of this kind as a promising approach to solve 
interesting 
problems in statistical physics. 

%The agreement found herein lead us to conclude that more general mechanism of self-duality breaking
%are worth to be investigated as they seem to be an efficient semi-analytical method to improve the
%already very satisfactory agreement with numerical simulations.

\vspace{0.5 cm}

\section*{Acknowledgements}

This work was supported by Fondecyt grant 3070055. The Centro de Estudios Cient\'{\i}ficos (CECS) is funded
by the Chilean Government through the Millennium Science Initiative and the
Centers of Excellence Base Financing Program of Conicyt. CECS is also
supported by a group of private companies which at present includes
Antofagasta Minerals, Arauco, Empresas CMPC, Indura, Naviera Ultragas and
Telef\'{o}nica del Sur. 
G.G. thanks the support of UBA, ANPCyT and CONICET, through UBACyT X861, PICT34577, PIP6160.

\end{document}